\documentclass[aps,prb,twocolumn,showpacs,showkeys,footinbib,groupedaddress]{revtex4-1}

\usepackage{amsmath}
\usepackage{amssymb}
\usepackage{graphicx}
\usepackage{bm}
\usepackage{color}
\usepackage{relsize}
\usepackage{braket}

\renewcommand{\i}{\ensuremath{\mathrm{i}}}
\newcommand{\e}{\ensuremath{\mathrm{e}}}

\usepackage{hyperref}

\begin{document}
\title{Probing Majorana-like states in quantum dots and quantum rings}
\author{Benedikt Scharf}
\affiliation{Department of Physics, University at Buffalo, State University of New York, Buffalo, NY 14260, USA}
\author{Igor \v{Z}uti\'c}
\affiliation{Department of Physics, University at Buffalo, State University of New York, Buffalo, NY 14260, USA}

\date{\today}

\begin{abstract}
Engineering chiral $p$-wave superconductivity in semiconductor structures offers fascinating ways to obtain and study Majorana modes in a condensed matter context. Here, we theoretically investigate chiral $p$-wave superconductivity in quantum dots and quantum rings. Using both analytical as well as numerical methods, we calculate the quasiparticle excitation spectra in these structures and the corresponding excitation amplitudes and charge densities. In the topological regime, we can observe the chiral edge modes localized at the boundaries and possessing finite energy in quantum dots and quantum rings. By applying a magnetic field which is expelled from the quantum ring, but which creates a flux that is an odd integer multiple of $\Phi_0/2=\pi\hbar/e$, Majorana modes, that is, (approximately) degenerate edge modes with zero energy and zero charge density, become possible in the topological regime. Furthermore, we investigate finite-size effects that split these degenerate edge modes as well as the effect of a magnetic field penetrating into the superconducting region that can under certain circumstances still support edge modes with approximately zero energy and charge.
\end{abstract}


\maketitle

\section{Introduction}\label{Sec:Intro}

Originally proposed almost eight decades ago, Majorana fermions are their own antiparticles, unlike, for example, electrons and their positronic counterparts.\cite{Majorana1937:NC} While in high-energy physics the concept of Majorana fermions remains important, albeit not experimentally demonstrated yet,\cite{Wilczek2009:NatPhys} the search for Majorana fermions in condensed-matter systems has developed into a topic of immense research interest in recent years.\cite{Franz2010:Physics,Alicea2012:RPP,Leijnse2012:SST,Beenakker2013:ARCMP,Franz2013:NatureNano,Tkachov2013:PSS} The attraction of pursuing Majorana fermions in solid-state systems is twofold: First, setups exhibiting Majorana modes could be more easily tailored in such systems. Secondly, Majorana fermions in a solid-state context are governed by non-Abelian statistics, which makes them potentially useful for topological quantum computation.\cite{Nayak2008:RMP,Alicea2011:NatPhys}

\begin{figure}[t]
\centering
\includegraphics*[width=8cm]{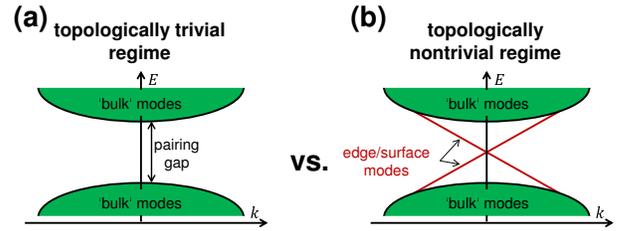}
\caption{(Color online) Excitation spectrum of a topological superconductor in the (a) trivial and (b) nontrivial regimes. By closing and subsequently reopening the pairing gap a transition between both regimes can be induced.}\label{fig:TS}
\end{figure}

In contrast to high-energy physics, where Majorana fermions are proposed to be fundamental particles, Majorana fermions in a solid-state context are emergent quasiparticle excitations. Since Majorana modes need to be their own antiparticles, superconductors offer a natural choice for systems in which to look for such excitations. The reason for this is that, in the Bardeen-Cooper-Schrieffer (BCS) theory of superconductivity, quasiparticle excitations are described by superpositions of both electrons and holes. Thus, excitations which are their own antiparticles, that is, excitations which are described by operators that are their own Hermitian conjugate, are possible at zero energy in certain types of superconductors. These superconductors, termed topological superconductors, possess a bulk pairing gap and---in their topologically nontrivial regime---gapless edge or surface states which can then under certain circumstances support Majorana fermions as midgap states (see Fig.~\ref{fig:TS}).\cite{Hasan2010:RMP,Qi2011:RMP,Shen2012}

In order to realize Majorana fermions in condensed-matter physics, it has long been proposed to utilize the midgap states of spinless chiral $p$-wave superconductors.\cite{Kopnin1991:PRB,Volovik1999:JETP,Senthil2000:PRB,Read2000:PRB,Kitaev2001:PhysUs,Sengupta2001:PRB,DasSarma2006:PRB} While some superconducting materials such as Sr$_2$RuO$_4$ have been argued to possess a superconducting state with $p$-wave symmetry,\cite{Mackenzie2003:RMP}\footnote{However, there are also possible alternative scenarios for chiral pairing symmetry [I. \v{Z}uti\'c and I. Mazin, Phys. Rev. Lett. {\bf 95}, 217004 (2005)]. For chiral $p$-wave symmetry, anomalous terms in the charge and current response have been proposed [R. M. Lutchyn, P. Nagornykh, and V. M. Yakovenko, Phys. Rev. B {\bf 77}, 144516 (2008)]. We do not consider such terms in this paper, nor a possible nonlinear relation between the superfluid velocity and the supercurrent  [K. Halterman and O. T. Valls, Phys. Rev. B {\bf 62}, 5904 (2000); I. \v{Z}uti\'c and O. T. Valls, Phys. Rev. B {\bf 58},  8738 (1998)].} it might be simpler to engineer $p$-wave pairing in more easily accessible systems. These proposals often, although not always, involve combining superconductivity, Zeeman splitting, and strong spin-orbit coupling,\cite{Sato2009:PRL,Sau2010:PRL,Sau2010:PRB,Alicea2010:PRB} which makes the field of spintronics\cite{Zutic2004:RMP,Fabian2007:APS} also relevant in this context. Typically, $p$-wave pairing is proximity-induced in materials with strong spin-orbit coupling, such as topological insulators\footnote{Proximity-induced superconductivity in topological insulators, however, is argued to require special attention [A. M. Black-Schaffer and A. V. Balatsky, Phys. Rev. B {\bf 86}, 144506 (2012)] and actually exhibits a mixed $s$- and $p$-wave character [G. Tkachov and E. M. Hankiewicz, Phys. Rev. B {\bf 88}, 075401 (2013)].} or semiconductors with strong Rashba coupling, by bringing them close to an $s$-wave superconductor. The Zeeman term then controls transitions between the topologically trivial and nontrivial regimes of the induced $p$-wave superconductor.

Proposals for such systems include the interface between a three-dimensional topological insulator and an $s$-wave superconductor, where Majorana modes can be bound to an Abrikosov vortex core,\cite{Fu2008:PRL} or wires, either semiconductor wires with strong Rashba spin-orbit coupling\cite{Lutchyn2010:PRL,Oreg2010:PRL,Klinovaja2012:PRL} or topological insulator wires,\cite{Cook2011:PRB,Cook2012:PRB} subject to a magnetic field and brought in close proximity to an $s$-wave superconductor. Besides, it has been suggested that Majorana modes can appear in systems that are in proximity to an $s$-wave superconductor, but do not require the presence of spin-orbit coupling.\cite{Kjaergaard2012:PRB,Braunecker2013:PRL,Vazifeh2013:PRL,Klinovaja2013:PRL} Conversely, $d$-wave superconductors\cite{Tsuei2000:RMP,Ngai2010:PRB} have been proposed as an alternative source for proximity-induced chiral $p$-wave pairing in Rashba or topological insulator structures with Zeeman splitting.\cite{Linder2010:PRL,Sato2010:PRB,Lucignano2012:PRB,Tsuei2013:Arxiv} Antiferromagnetically ordered chains of magnetic atoms on the surface of a conventional $s$-wave superconductor, where topologically unprotected Shiba states are converted into Majorana-bound states in a weak Zeeman field, offer another possible venue to Majorana modes.\cite{Heimes2014:PRB}

A setup consisting of two quantum dots with applied noncollinear magnetic fields and connected by an $s$-wave superconductor has been suggested to support modes localized in the dots that exhibit most properties of Majorana modes, but that are not topologically protected.\cite{Leijnse2012:PRB} Recent proposals to detect and control the Andreev reflection in topological insulator/superconductor/topological insulator\cite{Reinthaler2013:PRL} as well as in ferromagnet/superconductor\cite{Hoegl2015:Arxiv} junctions indicate a possibly similar control of proximity-induced topological superconductivity, crucial for the appearance of Majorana modes.

Potential signatures of Majorana modes in the above systems are usually centered on the fact that Majorana modes can also affect transport and thermodynamic properties of the (induced) topological superconductors.\cite{Beri2009:PRB,Flensberg2010:PRB,Wimmer2011:NJP,Lin2012:PRB,Tanaka2009:PRL,Law2009:PRL,Asano2010:PRL,Akhmerov2011:PRL,Schmeltzer2012:PRB,Tewari2012:PRB,Gibertini2013:PRB,Valentini2014:PRB,Grosfeld2011:PNAS,Sun2014:NJP,Kashuba2015:PRL,Valentini2015:PRB} Of all these proposals, the setup based on hybrid superconductor-semiconductor wire structures with Rashba spin-orbit coupling and Zeeman splitting\cite{Lutchyn2010:PRL,Oreg2010:PRL} has been the most prevalent until now, with experiments conducted in InSb\cite{Mourik2012:Science,Deng2012:NL,Rokhinson2012:NatPhys} and InAs\cite{Das2012:NatPhys,Finck2013:PRL} semiconductor wires as well as ferromagnetic atomic chains.\cite{NadjPerge2014:Science,Peng2015:PRL} In these experiments, a zero-bias peak in the tunneling conductance has been observed, which potentially points to the presence of Majorana zero-energy modes.\cite{Beri2009:PRB,Flensberg2010:PRB,Wimmer2011:NJP,Lin2012:PRB} However, mechanisms which could also give rise to this zero-bias conductance peak in the absence of Majorana modes, such as Kondo physics,\cite{Lee2012:PRL} strong disorder,\cite{Bagrets2012:PRL,Liu2012:PRL,Pikulin2012:NJP} smooth end confinement,\cite{Kells2012:PRB} or boundary effects,\cite{Roy2013:PRB} have been invoked. Moreover, experiments on semiconductor nanowire quantum dots strongly coupled to a conventional superconductor suggest that a zero-bias conductance anomaly with properties very similar to that expected for Majorana fermions can arise even without topological superconductivity.\cite{Lee2014:NatNano}

Thus, until now the experimental evidence for Majorana modes remains inconclusive,\cite{Prada2012:PRB,Rainis2013:PRB} although there are several proposals to supplement the tunneling conductance measurements in wires to verify the presence of Majorana-bound states or rule out some of the alternative sources for the zero-bias peak.\cite{DasSarma2012:PRB,Appelbaum2013:APL,Vernek2014:PRB,BenShach2015:PRB} Moreover, whereas disorder in chiral $p$-wave superconductor wires\cite{Rieder2013:PRB} is detrimental to the topological order supporting Majorana modes, disorder can even have a potentially stabilizing effect in the hybrid superconductor-semiconductor wire structures studied experimentally.\cite{Adagideli2014:PRB} Likewise, interaction effects in these hybrid structures are predicted to result in a reduction of the induced pairing gap,\cite{Gangadharaiah2011:PRL} but to actually expand the parameter range of the topologically nontrivial regime.\cite{Stoudenmire2011:PRB}

In contrast to these wire structures, we investigate the quasiparticle excitations, including possible Majorana modes, in quantum dots (QDs) and quantum rings (QRs) with (possibly induced) $p$-wave pairing. While there are several works on the ring geometry for the chiral $p$-wave superconductor and related models,\cite{Read2000:PRB,Stone2004:PRB,Fendley2007:PRB,Benjamin2010:PRB,Grosfeld2011:PNAS,Lucignano2013:PRB,Sun2014:NJP} our focus is on providing an in-depth analysis of the system and of the effects of finite size or magnetic fields penetrating into the superconducting region.

The paper is organized as follows: Section~\ref{Sec:Model} introduces the effective model used to describe the QDs and QRs. The results obtained from this model are then presented in Sec.~\ref{Sec:Results}, for QDs and QRs without any magnetic field (Sec.~\ref{SubSec:ZeroMagneticFlux}), QRs which enclose a magnetic flux, but which are not penetrated by a magnetic field (Sec.~\ref{SubSec:MagneticFlux}), and finally QRs into which a magnetic field can penetrate (Sec.~\ref{SubSec:MagneticField}). Section~\ref{Sec:Experiments} is devoted to the discussion of possible experimental realizations of the QDs and QRs as well as potential methods to distinguish Majorana modes from other phenomena. A brief summary in Sec.~\ref{Sec:Conclusions} concludes the paper.

\section{Model and methods}\label{Sec:Model}

\begin{figure}[t]
\centering
\includegraphics*[width=8cm]{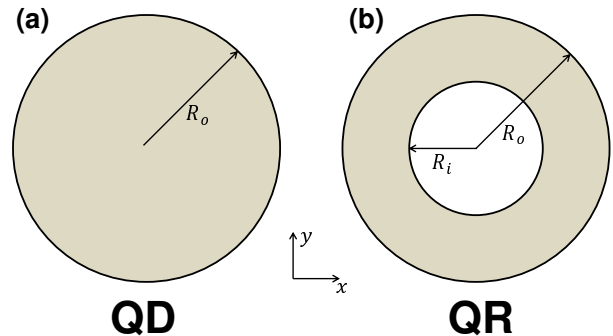}
\caption{(Color online) (a) QD and (b) QR considered in Sec.~\ref{SubSec:ZeroMagneticFlux}. The shaded areas are areas with $p$-wave superconductivity.}\label{fig:QDvsQR}
\end{figure}

In this paper, we consider QDs of radius $R_\mathrm{o}$ and QRs of inner radius $R_\mathrm{i}$ and outer radius $R_\mathrm{o}$ confined to the $xy$ plane and with chiral $p$-wave superconductivity (see Fig.~\ref{fig:QDvsQR}). Apart from certain superconducting materials such as Sr$_2$RuO$_4$, which has been argued to possess a state with $p$-wave symmetry, there are several proposals to engineer $p$-wave pairing in more easily accessible materials such as semiconductors with strong Rashba spin-orbit coupling by placing them close to an $s$-wave superconductor and exploiting proximity effects.\footnote{Rashba spin-orbit coupling at the interface of a normal-metal/$p$- or $d$-wave superconductor junction can also have a profound effect on the tunneling conductance [S.Wu and K. V. Samokhin, Phys. Rev. B {\bf 81}, 214506 (2010); S.Wu and K. V. Samokhin, Phys. Rev. B {\bf 82}, 184501 (2010)].} Since in the latter setup a magnetic field can also penetrate into the QDs or QRs without being expelled, we allow for the presence of a constant magnetic field $\mathbf{B}=B\mathbf{e}_z$ inside the QDs or QRs.

As a model for the interior of the QD and QR we consider a two-dimensional spinless $p$-wave superconductor, which can be described by the Bogoliubov-de Gennes (BdG) Hamiltonian\cite{Pientka2013:NJP}
\begin{equation}\label{BdGHamiltonian}
\begin{aligned}
\hat{H}=\left(\begin{array}{cc}
         H_0(\mathbf{r}) & \Delta(\mathbf{r}) \\
         \Delta^\dagger(\mathbf{r}) & -H_0^*(\mathbf{r}) \\
         \end{array}\right)
\end{aligned}
\end{equation}
in particle-hole space, where
\begin{equation}\label{pwaveHamiltonian}
H_0(\mathbf{r})=\frac{\left[\hat{\mathbf{p}}+e\mathbf{A}(\mathbf{r})\right]^2}{2m^*}-E_\mathrm{F}+V(r)
\end{equation}
is the single-particle Hamiltonian and
\begin{equation}\label{pwavePairing}
\Delta(\mathbf{r})=\alpha\e^{-\i n_\Phi\theta}\left\{\hat{p}_x-eA_x(\mathbf{r})-\i\left[\hat{p}_y-eA_y(\mathbf{r})\right]\right\}
\end{equation}
describes the pairing amplitude of the superconductor. Here, $\mathbf{r}$ denotes the position in the $xy$ plane, $\hat{\mathbf{p}}$ the momentum operator in the $xy$ plane, $\mathbf{A}(\mathbf{r})$ the magnetic vector potential in the $xy$ plane, $m^*$ the electronic effective mass, $E_\mathrm{F}$ the Fermi energy, $V(r)$ a radially symmetric confinement potential in the $xy$ plane, $\alpha=|\alpha|\e^{\i\varphi_\alpha}$ the $p$-wave pairing parameter, and $e=|e|$ the elementary charge. If a magnetic field is present, the $p$-wave pairing parameter acquires an additional phase of $-n_\Phi\theta$, where $\theta$ is the polar angular coordinate and $n_\Phi$ an integer related to the magnetic flux (see below). In the absence of a magnetic field, the sign of $E_\mathrm{F}$ determines whether the system described by Eq.~(\ref{BdGHamiltonian}) is in the topologically trivial regime ($E_\mathrm{F}<0$), where no edge modes appear, or in the topologically nontrivial regime ($E_\mathrm{F}>0$), where chiral edge modes can arise.

To model the confinement in the $xy$ plane, we use polar coordinates $(r,\theta)$ and infinite hard-wall potentials
\begin{equation}\label{confining_potential_QD}
V(r)=\left\{\begin{array}{ll}
            0 & \mathrm{for}\quad r<R_\mathrm{o}\\
	    \infty & \mathrm{elsewhere}
            \end{array}\right.
\end{equation}
and
\begin{equation}\label{confining_potential_QR}
V(r)=\left\{\begin{array}{ll}
            0 & \mathrm{for}\quad R_\mathrm{i}<r<R_\mathrm{o}\\
	    \infty & \mathrm{elsewhere}
            \end{array}\right.
\end{equation}
for QDs and QRs, respectively.

Moreover, we consider three different setups for the magnetic field: (i) no magnetic field, (ii) no magnetic field inside a superconductor QR, but with a magnetic flux $\Phi=\pi R^2_\mathrm{i}B$ enclosed by the QR, where the flux quantization of a superconductor requires $2\Phi/\Phi_0\in\mathbb{Z}$ with the magnetic flux quantum $\Phi_0=2\pi\hbar/e$, and (iii) a constant magnetic field $\mathbf{B}=B\mathbf{e}_z$ spread over the entire $xy$ plane and penetrating into the QR. For setups (i) and (ii), $n_\Phi=2\Phi/\Phi_0\in\mathbb{Z}$ and we choose the gauge $\mathbf{A}(\mathbf{r})=(\Phi/2\pi r)\mathbf{e}_\theta$, where $\mathbf{e}_\theta$ is the unit vector associated with the angular coordinate $\theta$ and $\Phi$ the flux enclosed by $R_\mathrm{i}$ [that is, $\Phi=0$ in setup (i) and $\Phi=BR^2_\mathrm{i}\pi$ in setup (ii)]. On the other hand, for setup (iii) we choose $\mathbf{A}(\mathbf{r})=(Br/2)\mathbf{e}_\theta$ and $n_\Phi=[2\Phi/\Phi_0]$ denotes the integer closest to $2\Phi/\Phi_0$, where $\Phi=\pi R^2_\mathrm{i}B$ is the magnetic flux enclosed by $r<R_\mathrm{i}$ in a QR, while in the case of a QD $n_\Phi=0$.

First, we note that for setups (i)-(iii) and a radially symmetric confinement $V(r)$ the commutator $\left[\hat{H},\hat{L}_\mathrm{eff}\right]=0$, where
\begin{equation}\label{effective_angular_momentum}
\hat{L}_\mathrm{eff}=\hat{L}_z+\frac{\hbar(1+n_\Phi)}{2}\tau_z
\end{equation}
is an effective angular momentum operator, $\hat{L}_z$ the angular momentum operator in $z$ direction, and $\tau_z$ the respective Pauli matrix in particle-hole space. Thus, the eigenstates of the BdG Hamiltonian~(\ref{BdGHamiltonian}) can be written as
\begin{equation}\label{ansatz}
\Psi(r)=\left(\begin{array}{c}
            u(\mathbf{r})\\
            v(\mathbf{r})\\
            \end{array}\right)=\frac{1}{\sqrt{2\pi}}\left(\begin{array}{c}
            \e^{\i m\theta}\,\e^{\i\varphi_\alpha/2}\,f(r)\\
            \e^{\i(m+1+n_\Phi)\theta}\,\e^{-\i\varphi_\alpha/2}\,g(r)\\
            \end{array}\right),
\end{equation}
where $\varphi_\alpha$ is the phase of $\alpha=|\alpha|\e^{\i\varphi_\alpha}$ and $m$ is the angular momentum of the electronic component, which we use as a quantum number to label the eigenstates.

Inserting this ansatz into the BdG equation leads to the radial equation
\begin{widetext}
\begin{equation}\label{radial_equation}
\left(\begin{array}{cc}
            -\frac{\hbar^2}{2m^*}\left\{\partial_r^2+\frac{1}{r}\partial_r-\frac{\left[m+n(r)\right]^2}{r^2}\right\}-E_\mathrm{F} & -\i\hbar|\alpha|\left[\partial_r+\frac{m+1+n_\Phi-n(r)}{r}\right]\\
            -\i\hbar|\alpha|\left[\partial_r-\frac{m+n(r)}{r}\right] & \frac{\hbar^2}{2m^*}\left\{\partial_r^2+\frac{1}{r}\partial_r-\frac{\left[m+1+n_\Phi-n(r)\right]^2}{r^2}\right\}+E_\mathrm{F}\\
            \end{array}\right)\left(\begin{array}{c}
            f(r)\\
            g(r)\\
            \end{array}\right)=E\left(\begin{array}{c}
            f(r)\\
            g(r)\\
            \end{array}\right)
\end{equation}
\end{widetext}
for $f(r)$ and $g(r)$ inside the QD ($r<R_\mathrm{o}$) or QR ($R_\mathrm{i}<r<R_\mathrm{o}$). In Eq.~(\ref{radial_equation}), $n(r)$ denotes the magnetic flux (per flux quantum) enclosed inside a disk of radius $r$, that is, $n(r)=n_\Phi/2$ for setups (i) and (ii) and $n(r)=Br^2\pi/\Phi_0$ for setup (iii). If energies and lengths are measured in terms of the Fermi energy $E_\mathrm{F}$ and the Fermi wavelength $\lambda_\mathrm{F}=2\pi\hbar/\sqrt{2m^*|E_\mathrm{F}|}$, the solutions depend only on $R_\mathrm{o}$, $R_\mathrm{i}$, $B$, and the effective pairing parameter
\begin{equation}\label{effective_parameter}
p=\frac{\hbar k_\mathrm{F}|\alpha|}{E_\mathrm{F}},
\end{equation} where the Fermi wavevector is given by $k_\mathrm{F}=2\pi/\lambda_\mathrm{F}$. In all three setups, the BdG Hamiltonian~(\ref{BdGHamiltonian}) exhibits particle-hole symmetry, that is, for each mode with energy $E$ there is another mode with energy -$E$. Consequently, for a given magnetic field or flux $\Phi$ the excitation energy denoted by the quantum number $m$, $E_m(\Phi)$, satisfies the relation $E_m(\Phi)=-E_{-(m+1+n_\Phi)}(\Phi)$.

To solve Eq.~(\ref{radial_equation}) and obtain the excitation spectrum and eigenstates, we employ a finite difference scheme. However, as detailed in Appendix~\ref{Sec:AppendixAnalyticalSolution}, we can also obtain the solutions for the setups (i) and (ii) analytically.

\section{Spectral and charge properties of quantum dots and quantum rings}\label{Sec:Results}

\subsection{No magnetic flux}\label{SubSec:ZeroMagneticFlux}

\begin{figure}[t]
\centering
\includegraphics*[width=8cm]{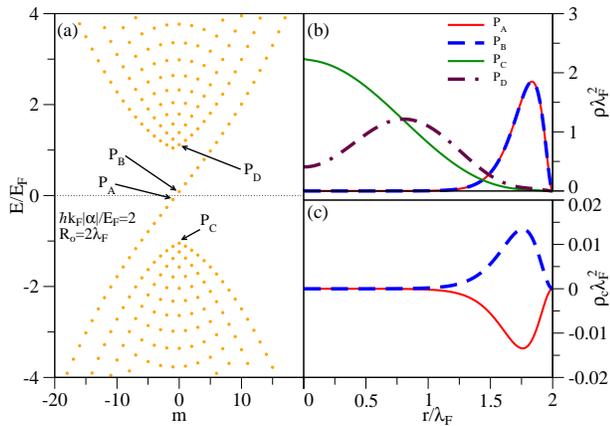}
\caption{(Color online) (a) Calculated excitation spectrum as well as (b) probability and (c) charge densities, $\rho(r)=\left|f(r)\right|^2+\left|g(r)\right|^2$ and $\rho_\mathrm{c}(r)=\left|f(r)\right|^2-\left|g(r)\right|^2$, respectively, for selected excitations in a QD with radius $R_\mathrm{o}=2\lambda_\mathrm{F}$ and pairing parameter $p=2$. Here, the states shown in panels~(b) and~(c) are marked in the energy spectrum, panel~(a). The charge densities of modes  $P_C$ and $P_D$ are not displayed in panel~(c) because their charge densities are two orders of magnitude larger than those of $P_A$ and $P_B$.}\label{fig:QD}
\end{figure}

\begin{figure}[t]
\centering
\includegraphics*[width=8cm]{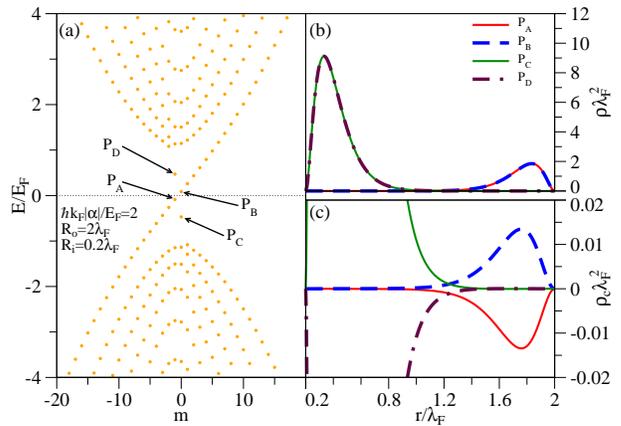}
\caption{(Color online) (a) Calculated excitation spectrum as well as (b) probability and (c) charge densities, $\rho(r)=\left|f(r)\right|^2+\left|g(r)\right|^2$ and $\rho_\mathrm{c}(r)=\left|f(r)\right|^2-\left|g(r)\right|^2$, respectively, for selected excitations in a QR with outer radius $R_\mathrm{o}=2\lambda_\mathrm{F}$, inner radius $R_\mathrm{i}=0.2\lambda_\mathrm{F}$, and pairing parameter $p=2$. Here, the states shown in panels~(b) and~(c) are marked in the energy spectrum, panel~(a).}\label{fig:QR}
\end{figure}

First, we investigate the situations of QDs and QRs where no magnetic field is present (see Fig.~\ref{fig:QDvsQR}), that is, case (i) where $n_\Phi=0$ and $n(r)=0$ in Eq.~(\ref{radial_equation}). Figures~\ref{fig:QD} and~\ref{fig:QR} show the excitation spectra as well as the amplitudes and charge densities of selected excitations for a QD and a QR in the topological regime ($E_\mathrm{F}>0$) with the pairing parameter $p=2$. The radius of the QD is chosen to be $R_\mathrm{o}=2\lambda_\mathrm{F}$, while the inner and outer radii of the QR are chosen as $R_\mathrm{i}=0.2\lambda_\mathrm{F}$ and $R_\mathrm{o}=2\lambda_\mathrm{F}$, respectively.

In Fig.~\ref{fig:QD}~(a), one can see that there is a gap in the excitation spectrum of the QD with the amplitudes of selected excitations ($P_C$, $P_D$) outside the gap shown in Fig.~\ref{fig:QD}~(b). As expected in the topological regime, however, there are also modes inside the gap ($P_A$, $P_B$), which correspond to charged modes localized at the edge of the QD [see Figs.~\ref{fig:QD}~(b) and~(c)]. A QR as illustrated in Fig.~\ref{fig:QR} possesses similar characteristics, that is, edge modes inside the superconducting gap ($P_A$, $P_B$), but now with additional charged modes (such as the modes $P_C$ and $P_D$ in Fig.\ref{fig:QR}) localized at the inner edge---at least as long as the spatial extent of the edge modes (essentially controlled by the parameter $p$ for small energies $E$, see also Sec.~\ref{SubSec:MagneticFlux} below as well as Appendix~\ref{Sec:AppendixAnalyticalSolution}) is small enough to prevent significant overlap between states localized at opposite edges.

However, in neither case, QD or QR, is there a mode with exactly zero energy (see below). Moreover, the energy spectrum in Fig.~\ref{fig:QR}~(a) illustrates that the slope of the inner edge modes is steeper than the slope of the outer edge modes. This can be qualitatively understood as a consequence of the conservation of the operator $\hat{L}_\mathrm{eff}$, which in turn requires eigenstates associated with a fixed electronic angular momentum quantum number $m$ to possess a higher group velocity if they are localized at the inner edge ($r\approx R_\mathrm{i}$) than if they are localized at the outer edge ($r\approx R_\mathrm{o}$).

\begin{figure}[t]
\centering
\includegraphics*[width=8cm]{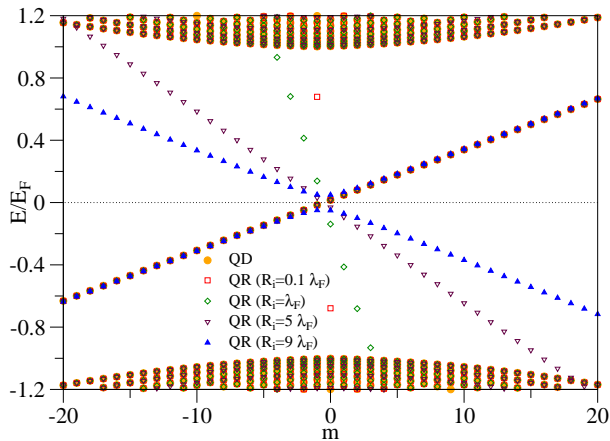}
\caption{(Color online) (a) Calculated excitation spectra for excitations in a QD and QRs with outer radius $R_\mathrm{o}=10\lambda_\mathrm{F}$, different inner radii $R_\mathrm{i}$, and pairing parameter $p=2$.}\label{fig:differentRi}
\end{figure}

Figure~\ref{fig:differentRi} shows how the energy spectrum of a QR with $p=2$ and an outer radius of $R_\mathrm{o}=10\lambda_\mathrm{F}$ depends on the inner radius $R_\mathrm{i}$. For comparison, we also show the energy spectrum of a QD with the same parameters as the QR. As $R_\mathrm{i}$ increases, there are two features that can be seen in Fig.~\ref{fig:differentRi}: First, the absolute value of the slope of the inner edge modes decreases with increasing $R_\mathrm{i}$, which can again be explained as originating from the conservation of the operator $\hat{L}_\mathrm{eff}$ and the resulting requirement that the group velocity should decrease as the radius increases. Second, the width $R_\mathrm{o}-R_\mathrm{i}$ of the QR decreases with increasing $R_\mathrm{i}$ and eventually the edge modes from the inner and outer edge modes overlap, which leads to the opening up of a hybridization gap as illustrated by the case of $R_\mathrm{i}=9\lambda_\mathrm{F}$ in Fig.~\ref{fig:differentRi}.

\begin{figure}[t]
\centering
\includegraphics*[width=8cm]{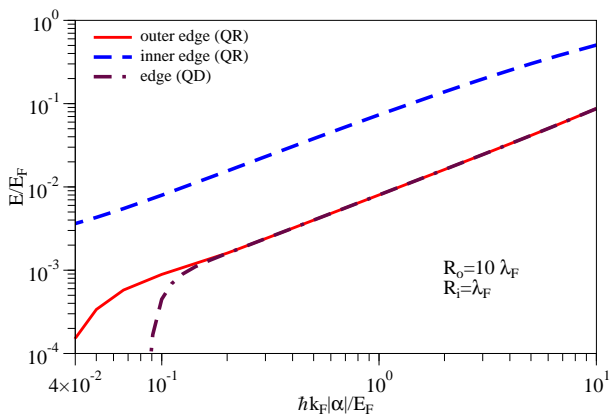}
\caption{(Color online) Dependence of the lowest energies of the edge modes in a QD with radius $R_\mathrm{o}=10\lambda_\mathrm{F}$ as well as of the inner and outer edge modes in a QR with radii $R_\mathrm{o}=10\lambda_\mathrm{F}$ and $R_\mathrm{i}=\lambda_\mathrm{F}$ as a function of the pairing parameter $p$.}\label{fig:QDQR_parameter}
\end{figure}

Next, we investigate the dependence of the lowest energy of the edge modes on the pairing parameter $p$, Eq.~(\ref{effective_parameter}), for QDs as well as for QRs. As illustrated by Fig.~\ref{fig:QDQR_parameter}, for large values of $p$ the lowest energy of the modes localized at the outer edge can be fitted to $E/E_\mathrm{F}\approx8.5\times10^{-3}p$ for both a QD and a QR. This agrees reasonably well to
\begin{equation}\label{Roi_approx}
\frac{E}{E_\mathrm{F}}=\frac{p}{4\pi(R_\mathrm{o/i}/\lambda_\mathrm{F})},
\end{equation}
as also found in Ref.~\onlinecite{Alicea2012:RPP} and which yields $E/E_\mathrm{F}\approx8.0\times10^{-3}p$ for $R_\mathrm{o}=10\lambda_\mathrm{F}$. Likewise, the lowest energy of the modes localized at the inner edge of a QR can be fitted to $E/E_\mathrm{F}\approx7.0\times10^{-2}p$, while Eq.~(\ref{Roi_approx}) yields $E/E_\mathrm{F}\approx8.0\times10^{-2}p$ for $R_\mathrm{i}=\lambda_\mathrm{F}$. The deviations from the behavior described by the approximation~(\ref{Roi_approx}) can be ascribed to the fact that in deriving this approximation curvature terms, that is, the kinetic terms in the diagonal elements of Eq.~(\ref{radial_equation}), have been neglected. For small values of $p$, curvature terms become even more important and the decrease of the lowest energy of the outer edge modes with decreasing $p$ is more pronounced. However, the energy is never exactly zero even for very small values of $p$. This situation, on the other hand, can change if a magnetic flux is induced in the QR, which will be discussed in the following section.

\subsection{Magnetic flux with no field penetration into the superconducting region}\label{SubSec:MagneticFlux}

\begin{figure}[t]
\centering
\includegraphics*[width=8cm]{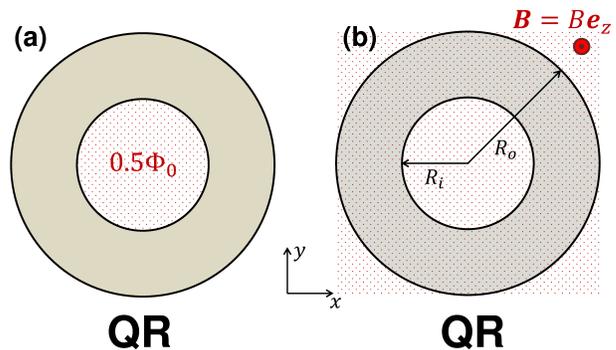}
\caption{(Color online) QRs as considered in Secs.~\ref{SubSec:MagneticFlux} and~\ref{SubSec:MagneticField}: (a) with an enclosed magnetic flux and no magnetic field penetrating into the QR and (b) a constant magnetic field. The shaded areas are areas with $p$-wave superconductivity.}\label{fig:QRBfield}
\end{figure}

In this and the following section, we study cases (ii) and (iii), that is, the effects of a perpendicular magnetic field on a QR as shown in Fig.~\ref{fig:QRBfield}. First, we look at situation (ii), where the magnetic field is completely expelled from the QR and just induces a magnetic flux enclosed by $R_\mathrm{i}$ as shown in Fig.~\ref{fig:QRBfield}~(a). As mentioned above, flux quantization in this case requires the flux to be $\Phi=n_\Phi\Phi_0/2$, where $n_\Phi$ is an integer. If $n_\Phi$ is even, the results for QRs from the previous section are recovered, but with the magnetic flux shifting states and energies denoted by the electronic angular momentum quantum number $m$ at zero flux to states denoted by $m-n_\Phi/2$, that is, $E_m(\Phi)=E_{m+n_\Phi/2}(0)$ [see also Eq.~(\ref{radial_equation_setups_i_ii}) in Appendix~\ref{Sec:AppendixAnalyticalSolution}].

While an even integer multiple of $\Phi_0/2$ thus essentially reduces the problem to the situation of zero magnetic flux, the situation is different for an odd integer multiple of $\Phi_0/2$. Without loss of generality we can choose $\Phi=\Phi_0/2$ in that case because every other odd integer multiple of $\Phi_0/2$ can be mapped to this flux, similar to how an even integer flux can be mapped to the situation of zero flux.

\begin{figure}[t]
\centering
\includegraphics*[width=8cm]{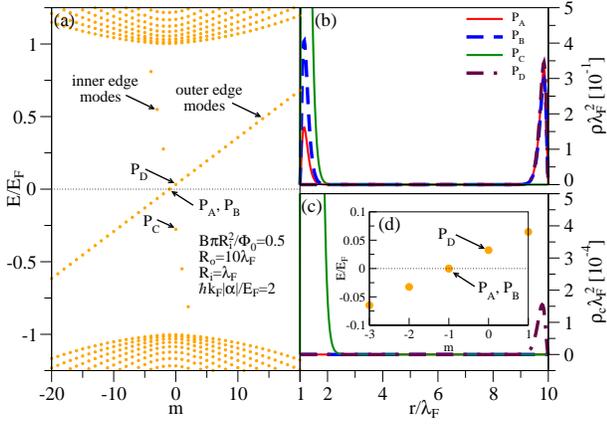}
\caption{(Color online) (a) Calculated excitation spectrum as well as (b) probability and (c) charge densities, $\rho(r)=\left|f(r)\right|^2+\left|g(r)\right|^2$ and $\rho_\mathrm{c}(r)=\left|f(r)\right|^2-\left|g(r)\right|^2$, respectively, for selected excitations in a QR with outer radius $R_\mathrm{o}=10\lambda_\mathrm{F}$, inner radius $R_\mathrm{i}=\lambda_\mathrm{F}$, and pairing parameter $p=2$ if a magnetic flux of $\Phi_0/2$ is enclosed by $R_\mathrm{i}$ with no magnetic field penetrating into the QR. Here, the states shown in panels~(b) and~(c) are marked in the energy spectrum, panel~(a). The amplitude of $P_D$ in panel~(b) is nearly completely overlapping with the two modes $P_A$ and $P_B$ at the outer radius. Panel~(d) displays the excitation energies close to zero.}\label{fig:QR_hif_oih}
\end{figure}

Figure~\ref{fig:QR_hif_oih} displays the situation for a QR in the topological regime ($E_\mathrm{F}>0$) if a magnetic flux of $\Phi_0/2$ is enclosed inside the QR and no magnetic field penetrates into the superconducting region. Here, we use the radii $R_\mathrm{i}=\lambda_\mathrm{F}$ and $R_\mathrm{o}=10\lambda_\mathrm{F}$ and the parameter $p=2$. Similar to the situation of zero flux, there are modes with energies inside the superconducting gap and localized at the boundaries of the QR (such as $P_C$, $P_D$). In contrast to the situation of zero flux, however, there are now also two degenerate edge modes (labeled as $P_A$, $P_B$) at zero energy [see the inset Fig.~\ref{fig:QR_hif_oih}~(d) and also Refs.~\onlinecite{Fendley2007:PRB,Alicea2012:RPP}] which are also chargeless and thus Majorana modes. This degeneracy at zero energy is a consequence of particle-hole symmetry, which for each mode with energy $E$ requires the existence of another mode with energy -$E$.

It is important to note, however, that if there is a finite overlap between the wave functions localized at the inner and outer radii, these degenerate Majorana modes are split similar to Fig.~\ref{fig:differentRi} in Sec.~\ref{SubSec:ZeroMagneticFlux}. This overlap is affected in two ways, namely by the width $R_\mathrm{o}-R_\mathrm{i}$ of the ring and by the spatial extent of the edge states.

\begin{figure}[t]
\centering
\includegraphics*[width=8cm]{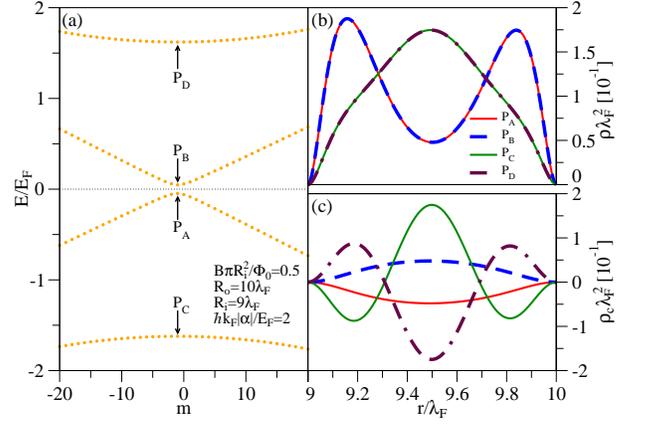}
\caption{(Color online) (a) Calculated excitation spectrum as well as (b) probability and (c) charge densities, $\rho(r)=\left|f(r)\right|^2+\left|g(r)\right|^2$ and $\rho_\mathrm{c}(r)=\left|f(r)\right|^2-\left|g(r)\right|^2$, respectively, for selected excitations in a QR with outer radius $R_\mathrm{o}=10\lambda_\mathrm{F}$, inner radius $R_\mathrm{i}=9\lambda_\mathrm{F}$, and pairing parameter $p=2$ if a magnetic flux of $\Phi_0/2$ is enclosed by $R_\mathrm{i}$ with no magnetic field penetrating into the QR. Here, the states shown in panels~(b) and~(c) are marked in the energy spectrum, panel~(a).}\label{fig:QR_hif_oihnarrow}
\end{figure}

The effect of a narrow width $R_\mathrm{o}-R_\mathrm{i}$ is illustrated in Fig.~\ref{fig:QR_hif_oihnarrow}, where we use a QR with the pairing parameter $p=2$, the flux $\Phi=\Phi_0/2$, and the inner and outer radii of $R_\mathrm{i}=9\lambda_\mathrm{F}$ and $R_\mathrm{o}=10\lambda_\mathrm{F}$, respectively. Here, the finite overlap between the two particle-hole symmetric modes $P_A$ and $P_B$ leads to a noticeable splitting between them of $E_{A/B}\approx\pm0.05E_\mathrm{F}$. This is due to a large overlap between the edge modes at the inner and the outer boundaries, which also results in a finite charge density of $P_A$ and $P_B$, similar to the situation in a wire geometry.\cite{Prada2012:PRB,DasSarma2012:PRB,Rainis2013:PRB,BenShach2015:PRB} In addition to the amplitudes and charge densities of $P_A$ and $P_B$, two excitations above the superconducting gap ($P_C$, $P_D$) are also displayed.

\begin{figure}[t]
\centering
\includegraphics*[width=8cm]{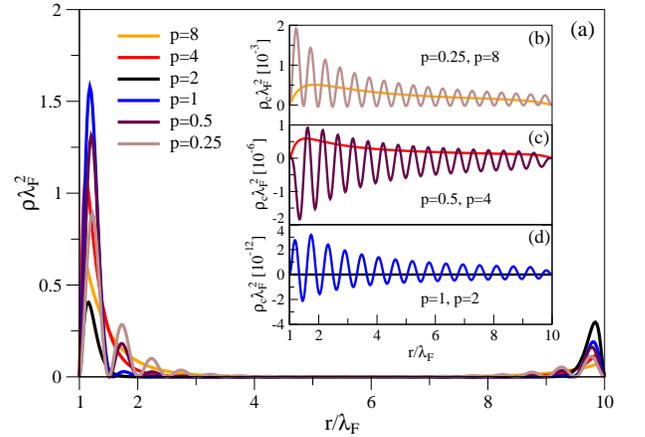}
\caption{(Color online) (a) Calculated probability and (b)-(d) charge densities, $\rho(r)=\left|f(r)\right|^2+\left|g(r)\right|^2$ and $\rho_\mathrm{c}(r)=\left|f(r)\right|^2-\left|g(r)\right|^2$, respectively, for the excitations at or close to zero energy in a QR with outer radius $R_\mathrm{o}=10\lambda_\mathrm{F}$, inner radius $R_\mathrm{i}=\lambda_\mathrm{F}$, and different pairing parameters $p$ if a magnetic flux of $\Phi_0/2$ is enclosed by $R_\mathrm{i}$ with no magnetic field penetrating into the QR. Panels~(b)-(d) demonstrate that for $p=2$ and $p=1$ the charge density is zero or extremely small, while other values of $p$ can lead to significantly higher charge densities.}\label{fig:QR_hif_oih_StatesForDifferentp}
\end{figure}

\begin{figure}[t]
\centering
\includegraphics*[width=8cm]{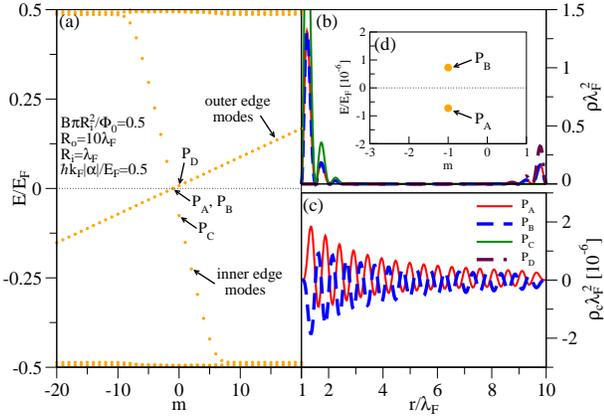}
\caption{(Color online) (a) Calculated excitation spectrum as well as (b) probability and (c) charge densities, $\rho(r)=\left|f(r)\right|^2+\left|g(r)\right|^2$ and $\rho_\mathrm{c}(r)=\left|f(r)\right|^2-\left|g(r)\right|^2$, respectively, for selected excitations in a QR with outer radius $R_\mathrm{o}=10\lambda_\mathrm{F}$, inner radius $R_\mathrm{i}=\lambda_\mathrm{F}$, and pairing parameter $p=0.5$ if a magnetic flux of $\Phi_0/2$ is enclosed by $R_\mathrm{i}$ with no magnetic field penetrating into the QR. Here, the states shown in panels~(b) and~(c) are marked in the energy spectrum, panel~(a). The charge densities of modes  $P_C$ and $P_D$ are not displayed in panel~(c) because their charge densities are four orders of magnitude larger than those of $P_A$ and $P_B$. Panel~(d) displays the excitation energies close to zero.}\label{fig:QR_hif_oihv2}
\end{figure}

A second parameter that affects the overlap between edge modes near zero energy is their spatial extent, which is essentially controlled by the pairing parameter $p$, Eq.~(\ref{effective_parameter}), where edge modes are most localized for values in the vicinity of $p=2$, while they become more extended if $p$ is decreased or increased [see Fig.~\ref{fig:QR_hif_oih_StatesForDifferentp}~(a)]. Likewise, Fig.~\ref{fig:QR_hif_oih_StatesForDifferentp}~(d) illustrates that close to $p=2$, where the charge density even vanishes and the excitation energy is zero, the charge density is very small as, for example, in the case of $p=1$. In Figs.~\ref{fig:QR_hif_oih_StatesForDifferentp}~(b) and~(c), on the other hand, the charge density increases by several orders of magnitude for values of $p$ farther away from $p=2$.\footnote{For open boundary conditions, the overlap between Majorana fermions is actually smallest for $p=\sqrt{2}$, where the lowest-order finite-difference scheme corresponds to the special case of the Kitaev chain from Ref.~\onlinecite{Kitaev2001:PhysUs} in which perfect Majorana fermions at exactly zero energy are localized at both ends of the chain.}

Moreover, Fig.~\ref{fig:QR_hif_oih_StatesForDifferentp} shows that for $p<2$ the probability and charge densities exhibit oscillations, while there is no oscillatory behavior for $p\geq2$. This behavior can be understood by looking at the form of the wave functions given by Eqs.~(\ref{radial_equation_fundamental_solutions}) and~(\ref{radial_equation_solution_kappa}) in Appendix~\ref{Sec:AppendixAnalyticalSolution}: The wave functions~(\ref{radial_equation_fundamental_solutions}) consist only of Bessel and Neumann functions, whose arguments are purely imaginary for $p\geq2$ and energies close to zero and thus result in non-oscillatory wave functions. For $p<2$, on the other hand, the arguments of the Bessel and Neumann functions are not purely imaginary for energies close to zero, which leads to oscillating wave functions.

To corroborate this statement, Fig.~\ref{fig:QR_hif_oihv2} shows the excitation spectrum, several amplitudes, and several charge densities of the same QR as in Fig.~\ref{fig:QR_hif_oih}, but with $p=0.5$. While in Fig.~\ref{fig:QR_hif_oih} we obtain two Majorana-bound states $P_A$ and $P_B$ as degenerate chargeless modes at exactly zero energy within the numerical accuracy of our calculation, there is a finite overlap between the two corresponding states $P_A$ and $P_B$ in Fig.~\ref{fig:QR_hif_oihv2}, which leads to a small, but finite splitting with $E_{A/B}\approx\pm10^{-8}E_\mathrm{F}$. Likewise, the charge carried by the particle-hole symmetric modes $P_A$ and $P_B$ in Fig.~\ref{fig:QR_hif_oihv2}~(c) is very small, but finite and spread over the entire ring [see also Figs.~\ref{fig:QR_hif_oih_StatesForDifferentp}~(b)-(d)], similar to the situation found in Ref.~\onlinecite{BenShach2015:PRB} for one-dimensional topological superconductor wires.\footnote{The agreement between the behavior of the charge density in one-dimensional wires detailed in Ref.~\onlinecite{BenShach2015:PRB} and in QRs is even more pronounced if the radial charge density $r\rho_\mathrm{c}(r)=r\left[\left|f(r)\right|^2-\left|g(r)\right|^2\right]$ is considered instead of $\rho_\mathrm{c}(r)$.}

In this sense, one can thus speak of the modes $P_A$ and $P_B$ in Fig.~\ref{fig:QR_hif_oihv2} as being Majorana-bound states only approximately. The difference between these modes and the remaining ordinary chiral edge modes away from zero energy can be revealed by their charge densities: While the charge density of those approximate Majorana-bound states is spread over the entire QR and very small as displayed in Figs.~\ref{fig:QR_hif_oih_StatesForDifferentp}~(b)-(d) and~\ref{fig:QR_hif_oihv2}~(c), the charge density of chiral edge modes away from zero energy, such as $P_C$ and $P_D$ in Fig.~\ref{fig:QR_hif_oih}~(c), is localized at the edges and typically several orders of magnitude larger.

\begin{figure}[t]
\centering
\includegraphics*[width=8cm]{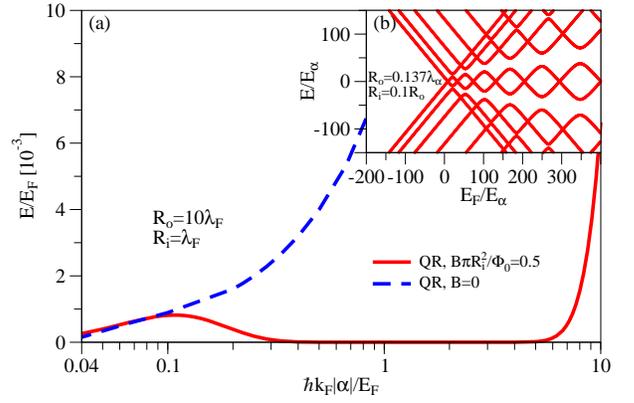}
\caption{(Color online) (a) Dependence of the lowest energies of the (outer) edge modes in a QR with radii $R_\mathrm{o}=10\lambda_\mathrm{F}$ and $R_\mathrm{i}=\lambda_\mathrm{F}$ as a function of the pairing parameter $p$ if no magnetic flux or a magnetic flux of $\Phi_0/2$ is enclosed by $R_\mathrm{i}$ with no magnetic field penetrating into the QR. The inset (b) shows the dependence of the energy spectrum at $m=-1$ on the Fermi energy for a QR which is threaded by a magnetic flux of $\Phi_0/2$. In the inset (b), energies are measured in units of $E_\alpha=m^*|\alpha|^2$ and lengths in units of $\lambda_\alpha=2\pi\hbar/m^*|\alpha|$.}\label{fig:Phi0p5lambdaFmudependence}
\end{figure}

The dependence of the lowest energy of the edge modes on the pairing parameter $p$ is displayed in Fig.~\ref{fig:Phi0p5lambdaFmudependence}~(a) for a QR with radii $R_\mathrm{o}=10\lambda_\mathrm{F}$ and $R_\mathrm{i}=\lambda_\mathrm{F}$ and a magnetic flux of $\Phi_0/2$. For comparison, the $p$ dependence in the case of zero flux is also included. While in the absence of any magnetic flux the energy increases monotonously as discussed in Fig.~\ref{fig:QDQR_parameter} in Sec.~\ref{SubSec:ZeroMagneticFlux}, the situation is different in the presence of a half-integer flux. Here, there is a region where the energy is very small and extremely close to zero, such as the situations depicted in Figs.~\ref{fig:QR_hif_oih} and~\ref{fig:QR_hif_oihv2}. As the value of $p$ is increased or decreased, for example, by changing the $p$-wave pairing $\alpha$ accordingly, the edge modes become more extended leading to a finite overlap and an increase in the splitting between the two particle-hole symmetric modes close to zero energy.

Figure~\ref{fig:Phi0p5lambdaFmudependence}~(b) shows the dependence of the energy spectrum at $m=-1$ on the Fermi energy of a QR threaded by a magnetic flux of $\Phi_0/2$ for a fixed $\alpha$. One can clearly see the transition between the topologically trivial regime for $E_\mathrm{F}<0$, where no edge modes with energies inside the bulk gap are present, and the topological regime, where such states arise. Moreover, those low-energy edge modes exhibit oscillations where the zeros are associated with jumps in the parity of the superconducting ground state.\cite{DasSarma2012:PRB,Rainis2013:PRB,BenShach2015:PRB} At these points, exact Majorana modes with zero energy and charge as in Fig.~\ref{fig:QR_hif_oih} can be found, while energies away from zero correspond to approximate Majorana modes like in Fig.~\ref{fig:QR_hif_oihv2}. With increasing distance $R_\mathrm{o}-R_\mathrm{i}$, the amplitude of these oscillations decreases as the overlap between edge modes situated at opposite edges decreases.

\begin{figure}[ht]
\centering
\includegraphics*[width=8cm]{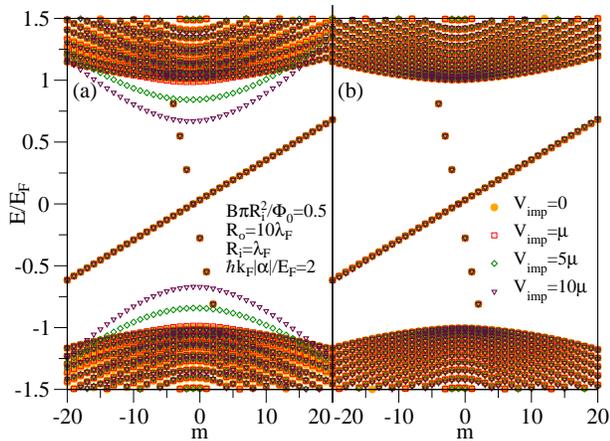}
\caption{(Color online) Calculated excitation spectra in a QR with outer radius $R_\mathrm{o}=10\lambda_\mathrm{F}$, inner radius $R_\mathrm{i}=\lambda_\mathrm{F}$, and pairing parameter $p=2$ if a magnetic flux of $\Phi_0/2$ is enclosed by $R_\mathrm{i}$ with no magnetic field penetrating into the QR and a concentric ring of an array of nonmagnetic impurities, characterized by the impurity potential $V_\mathrm{imp}$. The impurities are situated at $r=(R_\mathrm{o}-R_\mathrm{i})/2$ in panel~(a) and at $r=0.9775R_\mathrm{o}$ in panel~(b).}\label{fig:QR_impurity}
\end{figure}

Finally, using a very simple model, we can compare the effects of impurities on the edge and bulk modes. We consider a concentric ring of an array of nonmagnetic impurities characterized by the impurity potential $V_\mathrm{imp}$. Figure~\ref{fig:QR_impurity}~(a) shows the excitation spectrum for a QR with radii $R_\mathrm{o}=10\lambda_\mathrm{F}$ and $R_\mathrm{i}=\lambda_\mathrm{F}$, a magnetic flux of $\Phi_0/2$, and different strengths of $V_\mathrm{imp}$ for impurities situated at $r=(R_\mathrm{o}-R_\mathrm{i})/2$. Since the impurities are far away from the edges, the edge modes are not affected by the impurity potential, whereas the modifications of the bulk modes are clearly visible. On the other hand, if the impurities are situated at an edge, as is the case in Fig.~\ref{fig:QR_impurity}~(b), both the bulk and edge modes are not significantly changed. Thus, in contrast to bulk modes the edge modes are not substantially affected by nonmagnetic impurities, regardless of the position of the impurities. This striking difference between the role of perturbations on the edge and bulk states in Fig.~\ref{fig:QR_impurity} is consistent with the robustness that we can associate with the Majorana-like states.

\begin{figure}[t]
\centering
\includegraphics*[width=8cm]{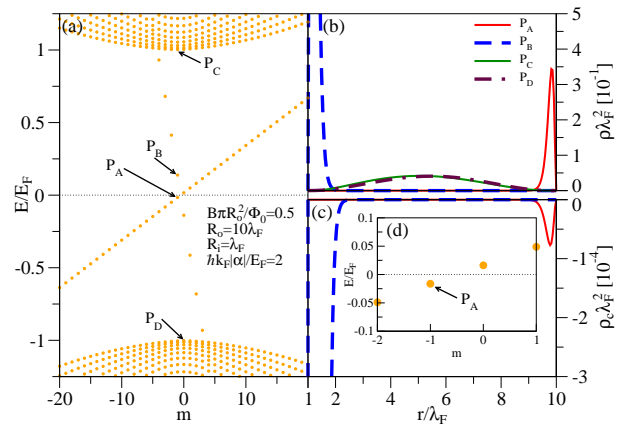}
\caption{(Color online) (a) Calculated excitation spectrum as well as (b) probability and (c) charge densities, $\rho(r)=\left|f(r)\right|^2+\left|g(r)\right|^2$ and $\rho_\mathrm{c}(r)=\left|f(r)\right|^2-\left|g(r)\right|^2$, respectively, for selected excitations in a QR with outer radius $R_\mathrm{o}=10\lambda_\mathrm{F}$, inner radius $R_\mathrm{i}=\lambda_\mathrm{F}$, and pairing parameter $p=2$ if a constant magnetic field penetrates into the QR with $R_\mathrm{o}$ enclosing a magnetic flux of $\Phi_0/2$. Here, the states shown in panels~(b) and~(c) are marked in the energy spectrum, panel~(a). Panel~(d) displays the excitation energies close to zero.}\label{fig:QR_hif}
\end{figure}

\subsection{Magnetic field penetrating into the superconducting region}\label{SubSec:MagneticField}

In the preceding discussions it has always been assumed that the magnetic field cannot penetrate into the superconducting region. Finally, we now also investigate the situation shown in Fig.~\ref{fig:QRBfield}~(b), where the perpendicular magnetic field can penetrate into the QR, that is, case (iii).

Our motivation for investigating this model is two-fold: As already mentioned, if the chiral $p$-wave BdG Hamiltonian in Eq.~(\ref{BdGHamiltonian}) is regarded as a simplified toy model for other, more complex hybrid structures with only proximity-induced pairing, a magnetic field can exist in these systems without being expelled like in a usual superconductor. On the other hand, if the system investigated is an extended three-dimensional superconductor, a magnetic field cannot penetrate far into this superconducting material and decays within the London penetration depth.\cite{DeGennes1989},\footnote{For unconventional pairing, components of the magnetic field may have a nonmonotonic dependence and even {\em increase} away from the surface of a superconductor, within the thickness of the London penetration depth [I. \v{Z}uti\'c and O. T. Valls, J. Comp. Phys. {\bf 136}, 337 (1997)].} However, if the sample is a thin film, that is, a film with a thickness smaller than the London penetration depth and a large extent of the base area, an applied in-plane magnetic field can be very accurately described as a constant field inside the film.\cite{Tinkham1996} Here, we make the simplifying assumption that such a behavior could also be applicable for an out-of-plane magnetic field.

Figure~\ref{fig:QR_hif} depicts a situation where a constant magnetic field $B$ spread over the entire $xy$ plane and corresponding to a magnetic flux $\pi R^2_\mathrm{o}B=0.5\Phi_0$ enclosed by the outer radius $R_\mathrm{o}$ is applied and thus induces no additional phase dependence of the $p$-wave pairing amplitude in Eq.~(\ref{BdGHamiltonian}), $n_\Phi=0$. The remaining parameters are chosen as $p=2$, $R_\mathrm{o}=10\lambda_\mathrm{F}$, and $R_\mathrm{i}=\lambda_\mathrm{F}$. Analyzing this setup, we can see that the behavior is very similar to the situation with zero flux, that is, there are edge modes with finite energies, but none with zero energy.

\begin{figure}[t]
\centering
\includegraphics*[width=8cm]{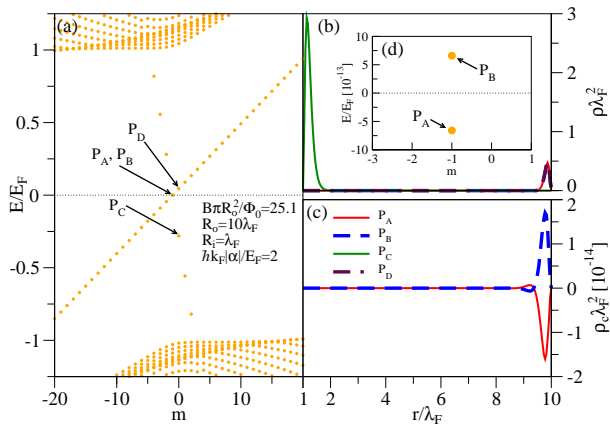}
\caption{(Color online) (a) Calculated excitation spectrum as well as (b) probability and (c) charge densities, $\rho(r)=\left|f(r)\right|^2+\left|g(r)\right|^2$ and $\rho_\mathrm{c}(r)=\left|f(r)\right|^2-\left|g(r)\right|^2$, respectively, for selected excitations in a QR with outer radius $R_\mathrm{o}=10\lambda_\mathrm{F}$, inner radius $R_\mathrm{i}=\lambda_\mathrm{F}$, and pairing parameter $p=2$ if a constant magnetic field penetrates into the QR with $R_\mathrm{o}$ enclosing a magnetic flux of $25.1\Phi_0$. Here, the states shown in panels~(b) and~(c) are marked in the energy spectrum, panel~(a). The amplitude of $P_D$ in panel~(b) is nearly completely overlapping with the two modes $P_A$ and $P_B$. The charge densities of modes  $P_C$ and $P_D$ are not displayed in panel~(c) because their charge densities are 14 orders of magnitude larger than those of $P_A$ and $P_B$. Panel~(d) displays the excitation energies close to zero.}\label{fig:QR_hif2}
\end{figure}

Next, we increase the strength of the magnetic field to a value corresponding to the magnetic flux $\pi R^2_\mathrm{o}B=25.1\Phi_0$, while keeping the remaining parameters $p$, $R_\mathrm{i}$, and $R_\mathrm{o}$ the same. For this strength of the magnetic field, an additional phase $-n_\Phi\theta$ with $n_\Phi=1$ is induced in the $p$-wave pairing amplitude, just like in case (ii) with a half-integer flux enclosed by $R_\mathrm{i}$, in which case Majorana modes---or more strictly speaking at least Majorana-like modes very close to zero energy and with tiny charge densities---were possible. Figure~\ref{fig:QR_hif2} illustrates that this is also the case here: The two particle-hole symmetric modes $P_A$ and $P_B$ display a finite splitting of $E_{A/B}\approx\pm7\times10^{-13}E_\mathrm{F}$ with extremely small, but finite charge densities, although in this case the charge densities are not spread over the QR, but localized at the edges [see Fig.~\ref{fig:QR_hif2}~(c)].

In this sense, Majorana-like modes (which originate from inducing a phase $-\theta$ in the $p$-wave pairing) persist even if magnetic fields penetrate into the superconducting region. For comparison, edge modes away from zero energy ($P_C$, $P_D$) are also shown in Fig.~\ref{fig:QR_hif2}. While the amplitude of the outer edge mode $P_D$ in Fig.~\ref{fig:QR_hif2}~(b) is nearly completely overlapping with the two Majorana-like modes $P_A$ and $P_B$, its charge density is 14 orders of magnitudes larger. Moreover, the energy spectrum outside the superconducting gap in Fig.~\ref{fig:QR_hif2}~(a) reflects the increasing importance of orbital effects as the strength of the magnetic field inside the superconducting ring increases.

\section{Experiment}\label{Sec:Experiments}

First, we briefly discuss possible systems in which the phenomena described above can be observed and which might be more readily available than intrinsic $p$-wave superconductor materials. Following several proposals,\cite{Sato2009:PRL,Sau2010:PRL,Sau2010:PRB,Alicea2010:PRB} a possible implementation of effective $p$-wave pairing can be in a semiconductor ring with strong Rashba spin-orbit coupling\cite{Zutic2004:RMP,Fabian2007:APS} that is brought into close proximity to an $s$-wave superconductor ring, which induces superconductivity into the semiconductor and which can be threaded by a magnetic flux $n_\Phi\Phi_0/2$. Flux quantization in a superconductor ring requires $n_\Phi$ to be an integer. Additionally, a Zeeman term is needed to break time-reversal symmetry and drive the induced $p$-wave superconductor into the topologically nontrivial phase.\footnote{Impurity-induced bound states offer a possibility to probe this topological superconductivity [H. Hu, L. Jiang, H. Pu, Y. Chen, and X.-J. Liu, Phys. Rev. Lett. {\bf 110}, 020401 (2013); J. D. Sau and E. Demler, Phys. Rev. B {\bf 88}, 205402 (2013)].} To engineer this, a ferromagnet is placed in the vicinity of the semiconductor ring. The ferromagnet then induces a Zeeman term $E_\mathrm{Z}$ into the semiconductor ring via the ferromagnetic proximity effect, while orbital terms induced by the ferromagnet can be omitted. Thus, the semiconductor ring is sandwiched between two rings, one superconducting the other ferromagnetic, which induce superconductivity and ferromagnetism, respectively, in the semiconductor ring.

Then, analogous to Eq.~(\ref{BdGHamiltonian}), the BdG Hamiltonian for the semiconductor ring (situated in the $xy$-plane) in particle-hole space is given by
\begin{equation}\label{BdGHamiltonianRealSystem}
\begin{aligned}
\hat{H}=\left(\begin{array}{cc}
         H_0(\mathbf{r}) & \Delta(\mathbf{r}) \\
         \Delta^\dagger(\mathbf{r}) & -\hat{s}_yH_0^*(\mathbf{r})\hat{s}_y \\
         \end{array}\right),
\end{aligned}
\end{equation}
 where
\begin{equation}\label{SingleParticleRealSystem}
\begin{aligned}
H_0(\mathbf{r})=&\frac{\left[\hat{\mathbf{p}}+e\mathbf{A}(\mathbf{r})\right]^2}{2m^*}-E_\mathrm{F}+V(r)\\
&+\alpha_\mathrm{SOC}\;\left\{\hat{\mathbf{s}}\times\left[\hat{\mathbf{p}}+e\mathbf{A}(\mathbf{r})\right]\right\}\cdot\mathbf{e}_z+E_\mathrm{Z}\;\mathbf{n}\cdot\hat{\mathbf{s}}
\end{aligned}
\end{equation}
and
\begin{equation}\label{PairingRealSystem}
\Delta(\mathbf{r})=\Delta\e^{-\i n_\Phi\theta}\hat{\mathbf{1}}
\end{equation}
are now matrices in spin-1/2 space. As before, $\mathbf{r}$, $\hat{\mathbf{p}}$, $\mathbf{A}(\mathbf{r})$, $m^*$, $E_\mathrm{F}$, and $V(r)$ denote the position in the $xy$ plane, the two-dimensional momentum operator, the magnetic vector potential, the electronic effective mass, the Fermi energy, and the confinement, while $\alpha_\mathrm{SOC}$ is the (Rashba) spin-orbit coupling strength, $\Delta$ the proximity-induced $s$-wave pairing amplitude,\footnote{As is usually done, we use a phenomenological $s$-wave pairing parameter $\Delta$ to model the superconducting proximity effect. A more sophisticated treatment, however, requires simultaneously solving the BdG equation and determining the pairing amplitude self-consistently [O. T. Valls, M. Bryan, and I. \v{Z}uti\'c, Phys. Rev. B {\bf 82}, 134534 (2010)]. Moreover, the semiconductor as potential Majorana host has to be in direct contact with the superconductor to induce localized Majorana fermions there [V. Stanev and V. Galitski, Phys. Rev. B {\bf 89}, 174521 (2014)].} $E_\mathrm{Z}$ the proximity-induced Zeeman term, $\hat{\mathbf{s}}$ a vector containing the three Pauli matrices in spin-1/2 space ($\hat{s}_x$, $\hat{s}_y$, and $\hat{s}_z$) and $\mathbf{n}$ the direction of the proximity induced magnetization.

\begin{figure}[t]
\centering
\includegraphics*[width=8cm]{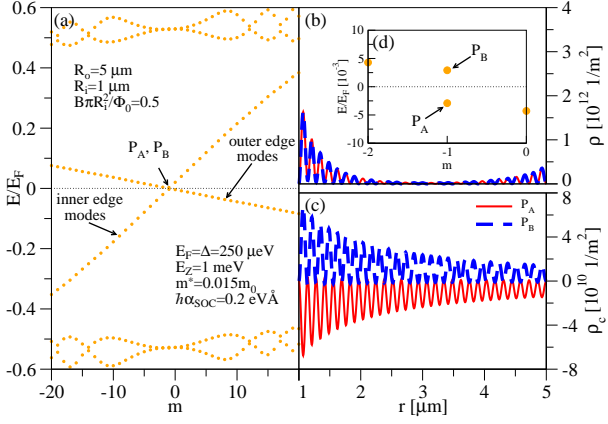}
\caption{(Color online) (a) Calculated excitation spectrum as well as (b) probability and (c) charge densities, $\rho(r)$ and $\rho_\mathrm{c}(r)$, respectively, of the approximate Majorana-bound states in an InSb ring with induced pairing and ferromagnetism, outer radius $R_\mathrm{o}=5$ $\mu$m, and inner radius $R_\mathrm{i}=1$ $\mu$m. Excitation energies close to zero, including the approximate Majorana-bound states (labeled $P_A$ and $P_B$) are shown in panel~(d).}\label{fig:QRRealSystem}
\end{figure}

The system described by Eqs.~(\ref{BdGHamiltonianRealSystem})-(\ref{PairingRealSystem}) exhibits the same features as the model of a spinless $p$-wave superconductor discussed in Secs.~\ref{Sec:Model} and~\ref{Sec:Results}, but with the condition $|E_\mathrm{Z}|-\sqrt{E^2_\mathrm{F}+|\Delta|^2}\lessgtr0$ separating the topological trivial ($<$) and nontrivial ($>$) regimes. Likewise, the conserved quantity is now given by the operator
\begin{equation}\label{effective_total_angular_momentum}
\hat{J}_\mathrm{eff}=\hat{L}_z+\frac{\hbar}{2}s_z+\frac{\hbar n_\Phi}{2}\tau_z
\end{equation}
instead of $\hat{L}_\mathrm{eff}$, that is, the orbital angular momentum $\hat{L}_z$ is replaced by the total angular momentum $\hat{L}_z+(\hbar/2)s_z$, while $n_\Phi+1$ is replaced by $n_\Phi$. Then, the eigenstates read as
\begin{equation}\label{StateRealSystem}
\Psi(r)=\frac{1}{\sqrt{2\pi}}\left(\begin{array}{c}
            \e^{\i m\theta}\,\e^{\i\varphi_\Delta/2}\,a(r)\\
            \e^{\i(m+1)\theta}\,\e^{\i\varphi_\Delta/2}\,b(r)\\
            \e^{\i (m+n_\Phi)\theta}\,\e^{-\i\varphi_\Delta/2}\,c(r)\\
            \e^{\i(m+1+n_\Phi)\theta}\,\e^{-\i\varphi_\Delta/2}\,d(r)\\
            \end{array}\right),
\end{equation}
where $\Delta=|\Delta|\e^{\i\varphi_\Delta}$. However, the basic conclusions from Secs.~\ref{Sec:Model} and~\ref{Sec:Results} are valid also for this system.

As an illustration of this, Fig.~\ref{fig:QRRealSystem} shows the excitation spectrum calculated from Eqs.~(\ref{BdGHamiltonianRealSystem})-(\ref{PairingRealSystem}) and corresponding to an InSb\cite{Mourik2012:Science} ($m^*=0.015m_0$, $\hbar\alpha_\mathrm{SOC}=0.2$ eV\AA) ring with an inner radius of $R_\mathrm{i}=1$ $\mu$m and an outer radius of $R_\mathrm{o}=5$ $\mu$m with an induced $s$-wave pairing amplitude of $\Delta=250$ $\mu$eV, an induced Zeeman splitting of $E_\mathrm{Z}=1$ meV in $z$-direction ($\mathbf{n}=\mathbf{e}_z$), and a Fermi energy of $E_\mathrm{F}=250$ $\mu$eV. Moreover, a magnetic flux of $\Phi=\Phi_0/2$ is enclosed by the rings without a magnetic field penetrating into the semiconductor, that is, we are looking at case (ii) from Secs.~\ref{Sec:Model} and~\ref{Sec:Results} with $n_\Phi=1$ and $\mathbf{A}(\mathbf{r})=(\Phi/2\pi r)\mathbf{e}_\theta=(\Phi_0/4\pi r)\mathbf{e}_\theta$.

For convenience and for the sake of easy comparison to the previous sections, we use the angular momentum $m$ of the electronic spin up component, that is, the first component in Eq.~(\ref{StateRealSystem}) to label the modes. Exactly as in the case of Fig.~\ref{fig:QR_hif_oihv2}, there are edge modes localized at the inner and outer radii as expected in the topological regime and two approximate Majorana-bound states $P_A$ and $P_B$ close to zero energy (with a splitting of $E_{A/B}\approx\pm7.3\times10^{-7}$ eV$\approx\pm2.9\times10^{-3}E_\mathrm{F}$) and with amplitudes and charge densities similar to those in Fig.~\ref{fig:QR_hif_oihv2}, as expected for a magnetic flux of $\Phi=\Phi_0/2$.\footnote{In a ring with an outer radius of $R_\mathrm{o}=10$ $\mu$m but with otherwise similar parameters to Fig.~\ref{fig:QRRealSystem}, the splitting would be $E_{A/B}\approx\pm2.7\times10^{-8}$ eV$\approx\pm1.0\times10^{-4}E_\mathrm{F}$.}

Following the proposal in Ref.~\onlinecite{BenShach2015:PRB} for a wire, where the situation is very similar, one possible way to experimentally detect the presence of Majorana modes in a QR could be by charge sensing: As shown in Secs.~\ref{SubSec:MagneticFlux}, the two Majorana-bound states are split into two particle-hole symmetric modes with very small, but finite energies if their wave functions localized at the inner and outer boundaries of the ring have a finite overlap. Likewise, those excitations possess a small, but finite charge density which is spread over the entire (radial extent of the) ring. This behavior is different from the other Andreev-bound states/chiral edge modes away from zero energy whose charge distribution is localized at a given edge. The charge distribution of a given mode at different points in the QR can then be probed by single-electron transistors acting as charge detectors.\cite{BenShach2015:PRB} In Appendix~\ref{Sec:QRWire_comparison}, we directly show the similarity of  Majorana-like states in a one-dimensional finite wire and a QR with cylindrical geometry.

Most proposals to detect Majorana modes in a condensed matter context usually involve conductance or---in the case of rings---interference measurements.\cite{Mourik2012:Science,Grosfeld2011:PNAS,Pientka2013:NJP,Sun2014:NJP} The signature of Majorana modes in these measurements is typically a zero-bias conductance peak or conductance peaks close to zero. However, these zero-bias conductance peaks can also originate from phenomena other than Majorana modes (such as the Kondo effect\cite{Lee2012:PRL}), which in turn makes it difficult to identify Majorana modes conclusively. Combining these conductance measurements with charge measurements as mentioned above could, on the other hand, then be used to identify the (approximate) Majorana modes.\cite{BenShach2015:PRB} This is because the charge distribution of split Majorana modes is spread over the entire ring, whereas phenomena such as the Kondo effect are restricted to the boundaries and possible changes in the charge density could only be detected there.

\section{Conclusions}\label{Sec:Conclusions}

We have theoretically studied chiral $p$-wave superconductivity in QDs and QRs and calculated the quasiparticle excitation spectra in these structures as well as the corresponding excitation amplitudes and charge densities. In the topological regime, we can observe the chiral edge modes localized at the boundaries and possessing finite energy in QDs and QRs, whereas no edge modes appear in the topologically trivial regime. However, none of the edge modes in the topological regime possess zero energy, that is, none of them is a Majorana mode. Only by applying a magnetic field which is expelled from the QR, but which creates a flux that is an odd integer multiple of $\Phi_0/2=\pi\hbar/e$, Majorana modes, that is, degenerate edge modes with zero energy and zero charge density, become possible in a QR in the topological regime, whereas none can be found in a QDs. Finite-size effects result in a splitting of these degenerate edge modes, leading to approximate Majorana modes in the sense that they have only approximately zero energy and zero charge density and are only approximately degenerate. This small, but finite charge distribution is then spread over the entire QR which allows for charge sensing measurements---in conjunction with other measurements at the edges---to probe the presence of Majorana modes. In the case of a magnetic field penetrating into the superconducting region, edge modes with approximately zero energy and charge can still be supported, although in this case the charge distribution is not necessarily spread over the entire QR.

\acknowledgments
We are grateful to \.{I}nan\c{c} Adagideli and Baris Pekerten for many discussions which were directly motivating this work. We thank Jong Han, Alex Matos-Abiague, and John Wei for stimulating discussions and suggestions. This work was supported by U.S. ONR Grant No. N000141310754 and DFG Grant No. SCHA 1899/1-1 (B.S.) as well as U.S. DOE, Office of Science BES, under Award DE-SC0004890 (I.\v{Z}.).

\appendix

\section{Analytical solution}
\label{Sec:AppendixAnalyticalSolution}

For setups (i) and (ii), Eq.~(\ref{radial_equation}) reduces to
\begin{widetext}
\begin{equation}\label{radial_equation_setups_i_ii}
\left(\begin{array}{cc}
            -\frac{\hbar^2}{2m^*}\left[\partial_r^2+\frac{1}{r}\partial_r-\frac{\left(m+n_\Phi/2\right)^2}{r^2}\right]-E_\mathrm{F} & -\i\hbar|\alpha|\left(\partial_r+\frac{m+1+n_\Phi/2}{r}\right)\\
            -\i\hbar|\alpha|\left(\partial_r-\frac{m+n_\Phi/2}{r}\right) & \frac{\hbar^2}{2m^*}\left[\partial_r^2+\frac{1}{r}\partial_r-\frac{\left(m+1+n_\Phi/2\right)^2}{r^2}\right]+E_\mathrm{F}\\
            \end{array}\right)\left(\begin{array}{c}
            f(r)\\
            g(r)\\
            \end{array}\right)=E\left(\begin{array}{c}
            f(r)\\
            g(r)\\
            \end{array}\right),
\end{equation}
\end{widetext}
where $m+n_\Phi/2$ is either an integer or a half-integer. We note that in the presence of a magnetic flux which is an integer multiple of $\Phi_0$, that is, in the case of $n_\Phi$ being an even integer, the energies $E$ and functions $f(r)$ and $g(r)$ for a fixed electronic angular momentum quantum number $m$ in Eq.~(\ref{radial_equation_setups_i_ii}) are determined by the same equation as the energies and functions for zero magnetic flux, but with an angular momentum quantum number $m+n_\Phi/2$. Thus, if $n_\Phi$ is an even integer, the energies and states with the angular momentum quantum number $m$ are the same as the energies and states with the angular momentum quantum number $m+n_\Phi/2$ at zero flux, that is, $E_m(\Phi)=E_{m+n_\Phi/2}(0)$.

After those remarks on the special case in which $n_\Phi$ is an even integer we now turn our attention to solving Eq.~(\ref{radial_equation_setups_i_ii}) for any integer $n_\Phi$. As mentioned in Sec.~\ref{Sec:Model}, the excitation energies $E_m(\Phi)$ satisfy the relation $E_m(\Phi)=-E_{-(m+1+n_\Phi)}(\Phi)$ due to particle-hole symmetry.

A solution to Eq.~(\ref{radial_equation_setups_i_ii}) can be obtained analytically because the diagonal components of the matrix correspond to Bessel's equation, while the off-diagonal elements correspond to raising and lowering operators for Bessel functions.\cite{Olver2010} A general solution to Eq.~(\ref{radial_equation_setups_i_ii}) at energy $E$ thus reads
\begin{equation}\label{radial_equation_solution}
\left(\begin{array}{ll}
            f(r)\\
            g(r)\\
            \end{array}\right)=a\eta_\mathsmaller{+}(r)+b\eta_\mathsmaller{-}(r)+c\chi_\mathsmaller{+}(r)+d\chi_\mathsmaller{-}(r)
\end{equation}
with the four fundamental solutions
\begin{equation}\label{radial_equation_fundamental_solutions}
\begin{aligned}
\eta_\mathsmaller{\pm}(r)=\left(\begin{array}{c}
            u_\mathsmaller{\pm}J_{l}(\kappa_\mathsmaller{\pm}r)\\
            v_\mathsmaller{\pm}J_{l+1}(\kappa_\mathsmaller{\pm}r)\\
            \end{array}\right),\quad
\chi_\mathsmaller{\pm}(r)=\left(\begin{array}{c}
            u_\mathsmaller{\pm}Y_{l}(\kappa_\mathsmaller{\pm}r)\\
            v_\mathsmaller{\pm}Y_{l+1}(\kappa_\mathsmaller{\pm}r)\\
            \end{array}\right),
\end{aligned}
\end{equation}
where $J_l(x)$ and $Y_l(x)$ are the Bessel and Neumann functions, $l=m+n_\Phi/2$ an integer or half-integer, 
\begin{equation}\label{radial_equation_solution_kappa}
\begin{aligned}
\kappa_\mathsmaller{\pm}=\kappa_\mathsmaller{\pm}(E)=&\frac{\sqrt{2m^*}}{\hbar}\bigg[E_\mathrm{F}-m^*|\alpha|^2\\
&\pm\sqrt{(E_\mathrm{F}-m^*|\alpha|^2)^2-E_\mathrm{F}^2+E^2}\bigg]^{1/2},
\end{aligned}
\end{equation}
and, if $\alpha\neq0$,
\begin{equation}\label{radial_equation_solution_AB}
\begin{aligned}
u_\mathsmaller{\pm}=\frac{\i\hbar|\alpha|\kappa_\mathsmaller{\pm}}{\sqrt{\hbar^2|\alpha|^2\kappa^2_\mathsmaller{\pm}+\left(\frac{\hbar^2\kappa^2_\mathsmaller{\pm}}{2m^*}-E_\mathrm{F}-E\right)^2}}\\
v_\mathsmaller{\pm}=\frac{\frac{\hbar^2\kappa^2_\mathsmaller{\pm}}{2m^*}-E_\mathrm{F}-E}{\sqrt{\hbar^2|\alpha|^2\kappa^2_\mathsmaller{\pm}+\left(\frac{\hbar^2\kappa^2_\mathsmaller{\pm}}{2m^*}-E_\mathrm{F}-E\right)^2}}.
\end{aligned}
\end{equation}
We note that the components $u_\mathsmaller{\pm}$ and $v_\mathsmaller{\pm}$ as well as $\kappa_\mathsmaller{\pm}$ (if measured in units of $k_\mathrm{F}$) depend only on the effective pairing parameter $p$ given by Eq.~(\ref{effective_parameter}) and the ratio $E/E_\mathrm{F}$. Thus, in the topological regime the spatial extent (with respect to $\lambda_\mathrm{F}$) of the edge modes at low energies is primarily determined by the parameter $p$. Finally, the energy $E$ and the complex coefficients $a$, $b$, $c$, and $d$ have to be determined from the boundary conditions for QDs and QRs.

\noindent\emph{(i) Quantum dots.}

In the case of QDs, the boundary conditions require that $f(r)$ and $g(r)$ do not diverge at $r=0$ and that $f(R_\mathrm{o})=g(R_\mathrm{o})=0$. The condition at $r=0$ can only be satisfied if $c=d=0$ in Eq.~(\ref{radial_equation_solution}), while the condition at $r=R_\mathrm{o}$ yields a system of two linear equations for the coefficients $a$ and $b$. This system has a nontrivial solution if
\begin{equation}\label{te_QD}
u_\mathsmaller{+}v_\mathsmaller{-}J_{l}(\kappa_\mathsmaller{+}R_\mathrm{o})J_{l+1}(\kappa_\mathsmaller{-}R_\mathrm{o})=u_\mathsmaller{-}v_\mathsmaller{+}J_{l}(\kappa_\mathsmaller{-}R_\mathrm{o})J_{l+1}(\kappa_\mathsmaller{+}R_\mathrm{o}),
\end{equation}
which in turn represents a transcendental equation to obtain the excitation energies $E$ for QDs.

\noindent\emph{(ii) Quantum rings.}

For QRs, the boundary conditions $f(R_\mathrm{o})=g(R_\mathrm{o})=f(R_\mathrm{i})=g(R_\mathrm{i})=0$ have to be satisfied. Inserting these conditions into Eq.~(\ref{radial_equation_solution}) yields a linear system of four equations for the coefficients $a$, $b$, $c$, and $d$, which possesses a nontrivial solution if
\begin{equation}\label{te_QR}
\det\left(\begin{array}{ccccc}\eta_\mathsmaller{+}(R_\mathrm{o}) & \eta_\mathsmaller{-}(R_\mathrm{o}) & & \chi_\mathsmaller{+}(R_\mathrm{o}) & \chi_\mathsmaller{-}(R_\mathrm{o})\\
\\
\eta_\mathsmaller{+}(R_\mathrm{i}) & \eta_\mathsmaller{-}(R_\mathrm{i}) & & \chi_\mathsmaller{+}(R_\mathrm{i}) & \chi_\mathsmaller{-}(R_\mathrm{i})
\end{array}\right)=0.
\end{equation}
The excitation energies $E$ for QRs can then be obtained from the transcendental Eq.~(\ref{te_QR}).

\section{Comparison between a quantum ring and a wire}
\label{Sec:QRWire_comparison}

\begin{figure}[t]
\centering{\includegraphics*[width=8cm]{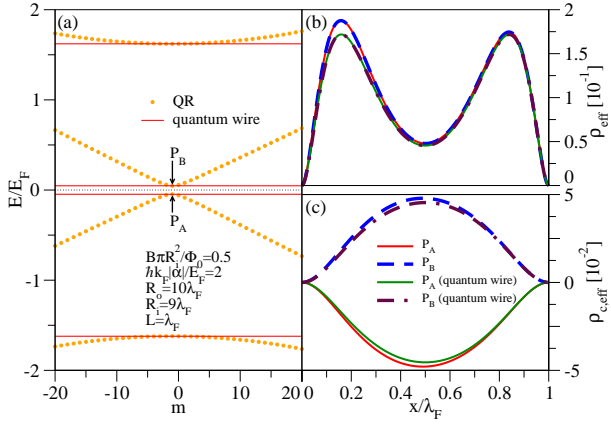}}
\caption{(Color online) Comparison between the (a) calculated excitation spectra as well as the effective dimensionless (b) probability and (c) charge densities, $\rho_\mathrm{eff}(x)$ and $\rho_\mathrm{c,eff}(x)$, respectively, for selected excitations in a QR threaded by a magnetic flux of $\Phi_0/2$ and a one-dimensional wire with the same parameters. While the energy levels in the QR can be described by the quantum number $m$, the quantum wire has discrete energy levels. Here, the states shown in panels~(b) and~(c) are marked in the energy spectrum, panel~(a).}\label{fig:QRWire_comparison}
\end{figure}

Figure~\ref{fig:QRWire_comparison} compares the excitation spectrum, probability and charge densities for selected excitations in a QR with outer radius $R_\mathrm{o}=10\lambda_\mathrm{F}$, inner radius $R_\mathrm{i}=9\lambda_\mathrm{F}$, and pairing parameter $p=2$ if a magnetic flux of $\Phi_0/2$ is enclosed by $R_\mathrm{i}$ with no magnetic field penetrating into the QR and a one-dimensional wire of length $L=\lambda_\mathrm{F}$, pairing parameter $p=2$, and no magnetic field. The coordinate $x$ in Figs.~\ref{fig:QRWire_comparison}~(b) and~(c) is to be read as the shifted radial coordinate $r-R_\mathrm{i}$ in a QR and the one-dimensional coordinate $l$ of a wire, while the effective dimensionless probability and charge densities are to be read as $\rho_\mathrm{eff}(x)=\rho(r)\lambda_\mathrm{F}^2$ and $\rho_\mathrm{c,eff}(x)=\rho_\mathrm{c}(r)\lambda_\mathrm{F}^2$ for the QR and as $\rho_\mathrm{eff}(x)=\rho(l)\lambda_\mathrm{F}$ and $\rho_\mathrm{c,eff}(x)=\rho_\mathrm{c}(l)\lambda_\mathrm{F}$ for the wire.

As can be seen in Fig.~\ref{fig:QRWire_comparison}, the eigenenergies of the QR at $m=-1$ correspond very well to the eigenenergies of a wire with the same parameters. Likewise, the probability and charge densities of the two geometries are very similar with a slight difference in their probability maxima in Fig.~\ref{fig:QRWire_comparison}~(b), which are different for the QR due to the inequivalent inner and outer edge modes, unlike for the wire geometry. The similarity is even more pronounced if instead of the densities $\rho(r)\lambda_\mathrm{F}^2$ and $\rho_\mathrm{c}(r)\lambda_\mathrm{F}^2$ the radial densities $\rho(r)r\lambda_\mathrm{F}$ and $\rho_\mathrm{c}(r)r\lambda_\mathrm{F}$ are chosen. Then, both maxima in the probability density of the QR have the same height similar to the wire.

\bibliography{BibTopInsAndTopSup}

\end{document}